          [2001/04/25 1.1 (PWD)]
\documentclass[a4paper,twocolumn]{esapub} % European paper
\usepackage{natbib}
\usepackage{graphicx}

\title{Predictions for Cosmological Infrared Surveys with Herschel from a Mid-Infrared 
Phenomenological Evolution Model}
\author{Carlotta Gruppioni}
\affil{INAF -- Osservatorio Astronomico di Bologna, via Ranzani 1, I--40127 Bologna, Italy}
\author{Francesca Pozzi}
\affil{Dipartimento di Astronomia, Universit\`a di Bologna, via Ranzani 1, I--40127 Bologna, Italy}
\author{Carlo Lari}
\affil{IRA--CNR, via Gobetti 101, I--40129 Bologna, Italy}

\begin{document}


\maketitle

\begin{abstract}
We make predictions for the cosmological surveys to be conducted by the future 
Herschel mission operating in the far-infrared. The far-infrared bands match 
the peak of the CIRB, the brightest background of astrophysical origin. 
Therefore, surveys in these bands will provide essential information on the 
evolutionary properties of Luminous and Ultra-Luminous Infrared Galaxies 
(LIGs and ULIGs), starburst and normal galaxies.
Our predictions are based on a new phenomenological model obtained from the 
15-$\mu$m luminosity function of 
galaxies and AGN, fitting all the ISOCAM observables (source counts and redshift
distributions) and also the recently published Spitzer source counts in the 
24-$\mu$m band.
We discuss the confusion noise due to extragalactic sources, depending strongly 
on the shape of the source counts and on the telescope parameters. 
We derive the fraction of the CIRB expected to be resolved by Herschel in the 
different wavebands and we discuss extragalactic surveys that could be carried 
on by Herschel for different scientific puropouses (i.e. ultra-deep, deep and shallow).
\end{abstract}

\section{Introduction}
The mid- and far-infrared (MIR and FIR) regions of the electromagnetic spectrum probe the 
population of actively star-forming galaxies obscured by dust.
Extragalactic source counts from different surveys over a wide flux range obtained with 
ISOCAM indicate that these sources have evolved rapidly, significantly faster than deduced 
from optical surveys (Elbaz et al. 1999; Gruppioni  et al. 2002). These results are supported 
by the detection of a substantial cosmic infrared background (CIRB, Hauser \& Dwek 2001), 
which is interpreted as the integrated emission from dust present in galaxies. 
The Spitzer Space Telescope is now providing new insights into the IR population contributing 
to the CIRB, in particular with the  MIPS 24-$\mu$m band, that is starting to detect a 
population of galaxies that may be IR luminous galaxies at z $\sim$ 1.5-3 (i.e. high-redshift 
analogs of the faint 15-$\mu$m galaxies detected by ISOCAM). The availability of the new 
space facilities in the coming years, such  as Herschel, and on the ground ALMA, opens new 
perspective to study in detail the population of IR  galaxies beyond z = 1.  It is now 
necessary to use models to make predictions for future surveys, in order to better plan the 
observations with these new facilities. In particular, we are now using models to answer 
to the following questions: Which galaxy populations will the new instruments detect?
What fraction of the CIRB will be detected into sources? Will we detect any evolution and up 
to what redshift? What will limit the future surveys? (confusion? sensitivity?)

Here we present the predictions for future Surveys with Herschel obtained with a new model based 
on the first determination of the 15-$\mu$m Luminosity Function, reproducing all the ISOCAM 
surveys observables and also the recent Spitzer 24-$\mu$m source counts (see F. Pozzi's paper).

\section{The Model}
We use a model fitting the observed 15-$\mu$m source counts (Pozzi et al. 2004; Matute et
al. in preparation) to make
predictions in the Herschel bands. The model is based on the first determination of the
15-$\mu$m luminosity function of galaxies and AGN, from data of the ELAIS southern fields
survey (Lari et al. 2001; La Franca et al. 2004; Rowan-Robinson et al. 2004). Four main
populations, evolving independently, contribute to the observed source counts: ``starburst''
and ``normal'' galaxies and type 1 and 2 AGN. The ``normal'' population is consistent with
no evolution, while the ``starburst'' population requires a strong evolution, both
in luminosity and density ($L(z) \propto (1+z)^{3.5}$ and $\rho(z) \propto (1+z)^{3.8}$
respectively, up to $z \sim 1$). 
\begin{figure*}
\centering
\vspace{-2cm}
\includegraphics[angle=0,width=0.45\linewidth]{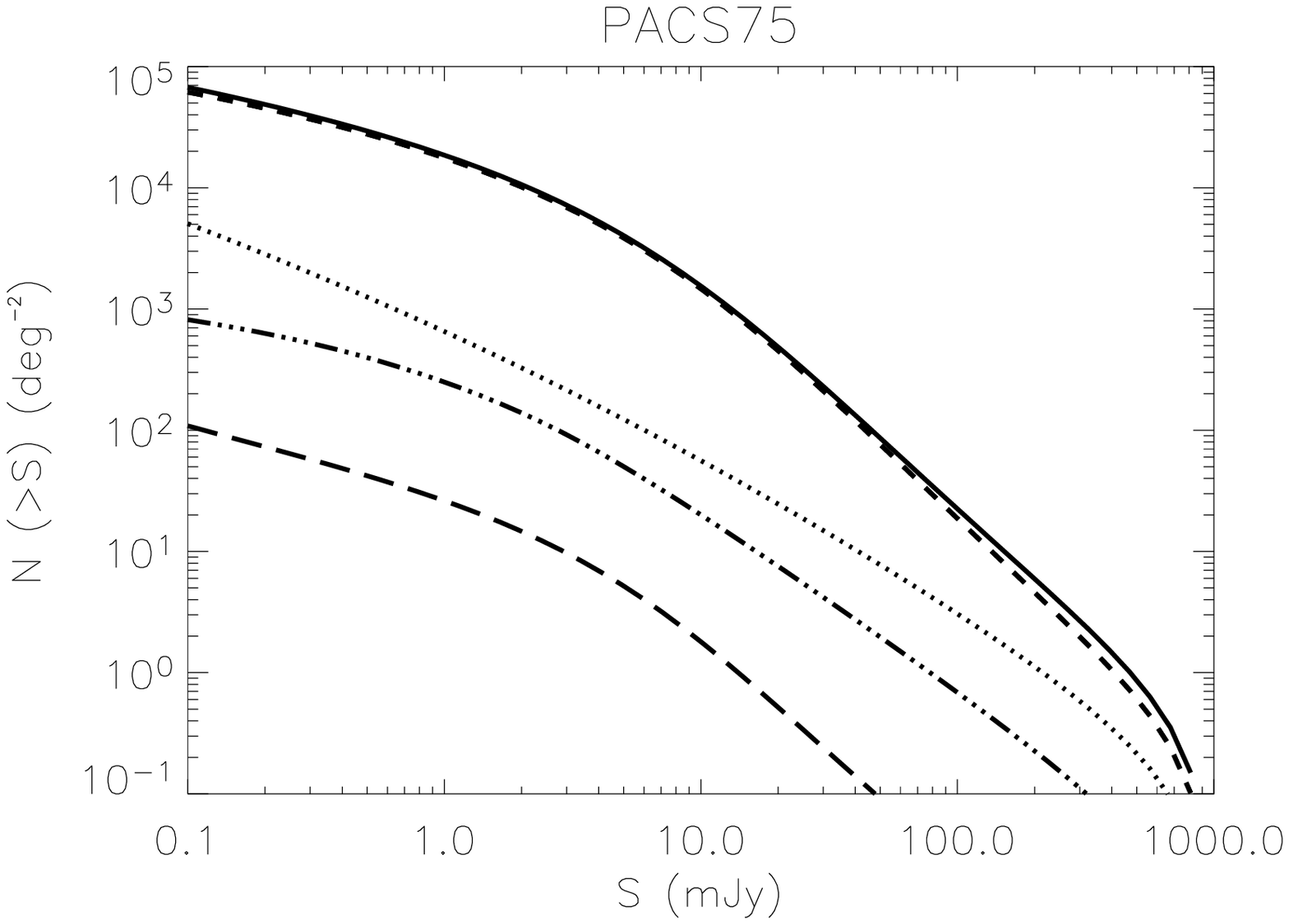}
\includegraphics[angle=0,width=0.45\linewidth]{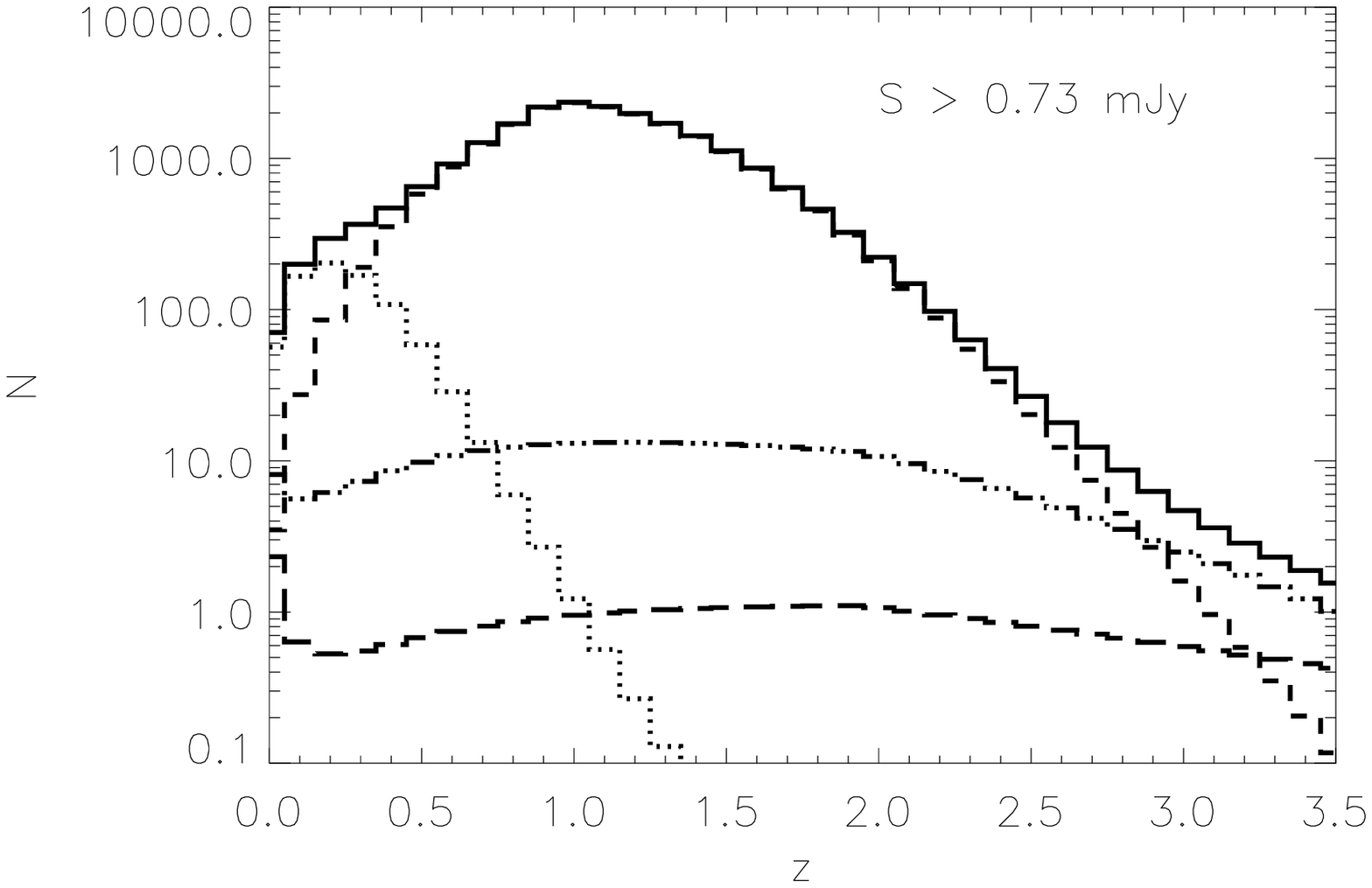}
\includegraphics[angle=0,width=0.45\linewidth]{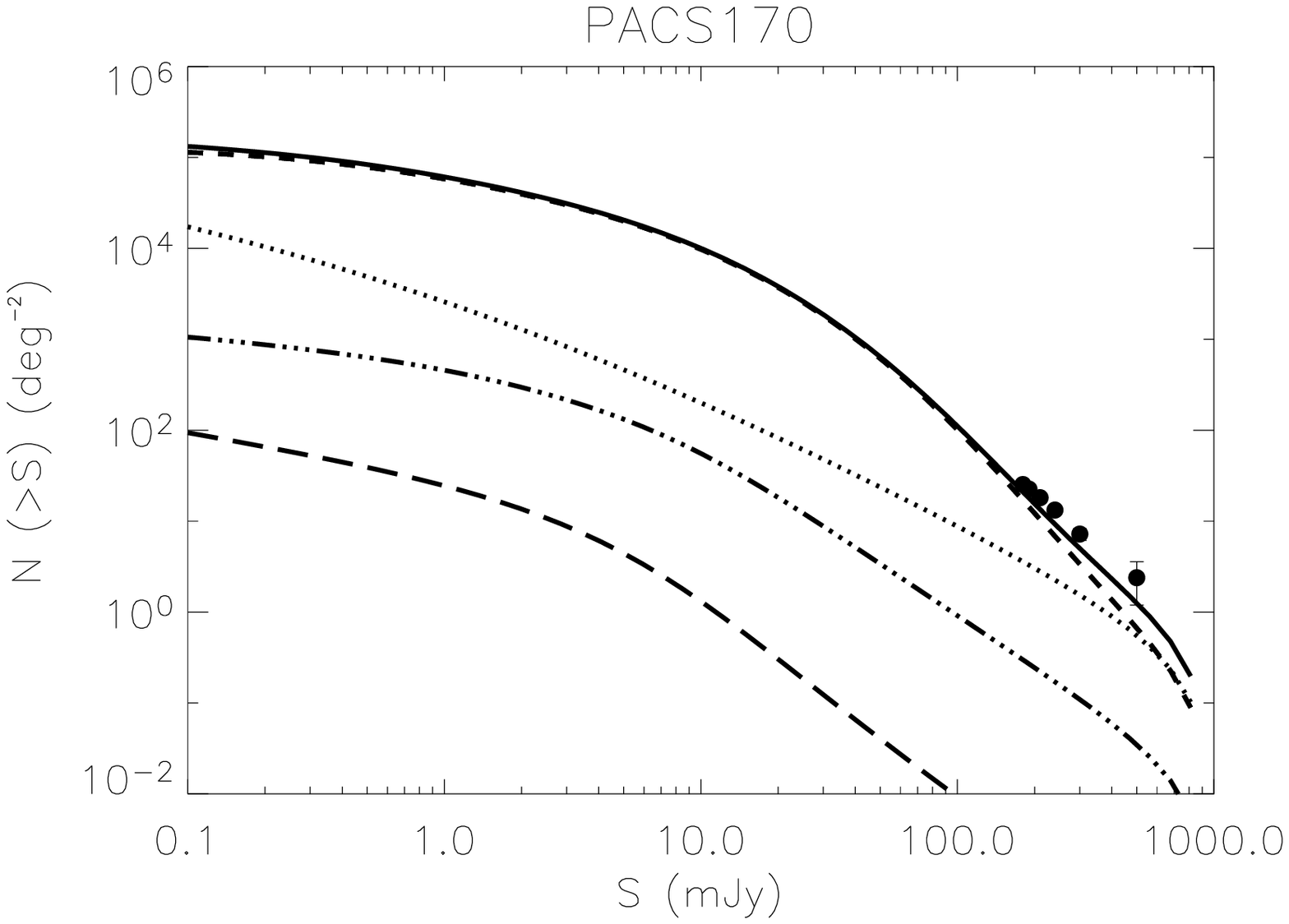}
\includegraphics[angle=0,width=0.45\linewidth]{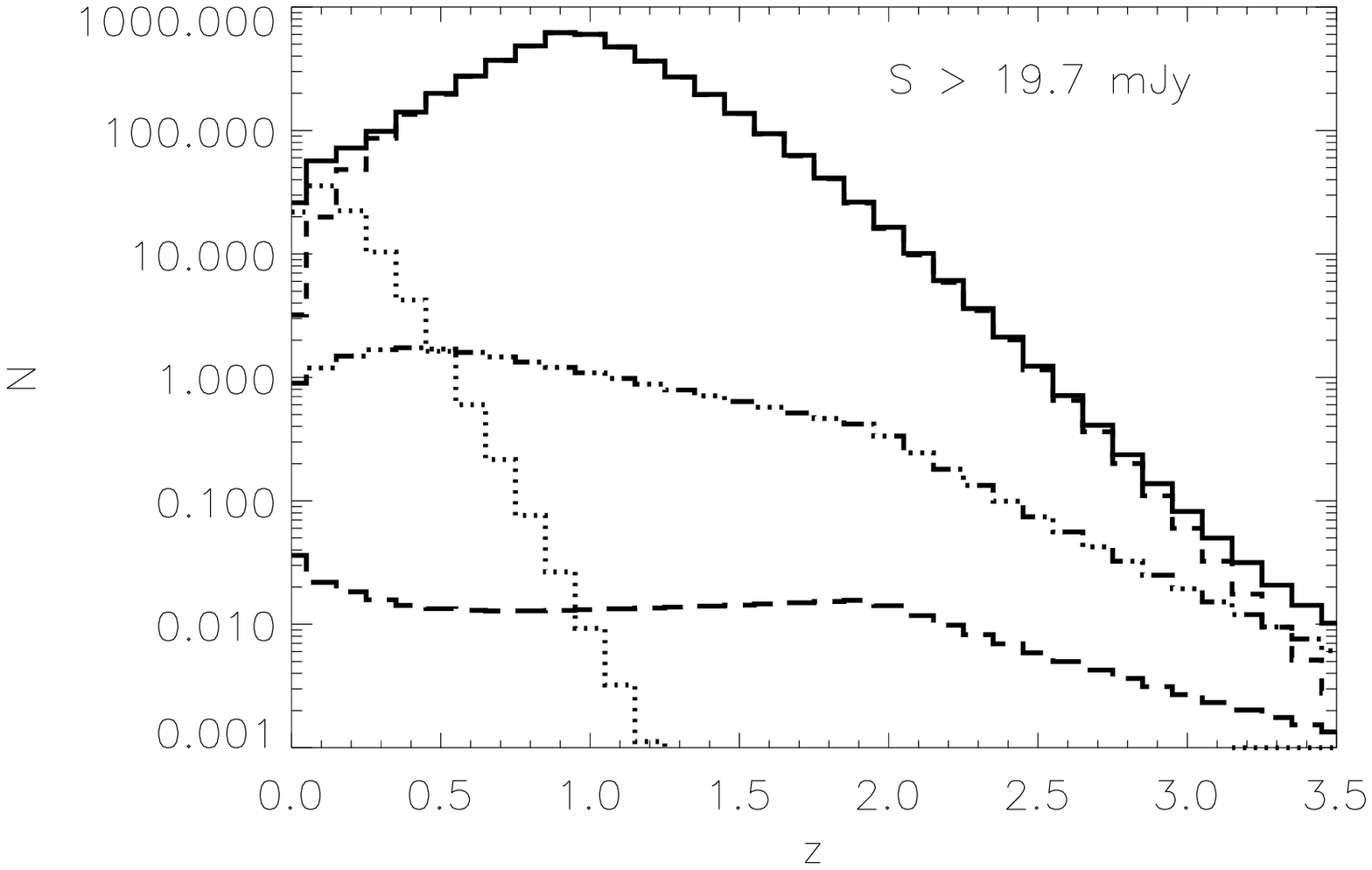}
\caption{Predictions in the 75-$\mu$m ($top$) and 170-$\mu$m ($bottom$) bands of PACS.
The predicted source counts and redshift distributions for the different source populations 
are shown in the $left$ and $right$ panels respectively. The dashed, dotted, long-dashed,
dot-dot-dot-dashed and solid lines are the predicted contribution of the ``starburst'', ``normal'',
agn1, agn2 and all populations respectively. The data points overplotted to the model counts
at 170 $\mu$m are the observed FIRBACK counts from Dole et al. (2001).\label{fig_cnt}}
\end{figure*}
The results of the model are in remarkable good
agreement with all the available 15-$\mu$m observables, like source counts at all flux
density levels (0.1 -- 300 mJy) and redshift distributions and luminosity functions
at low and high $z$ (see Pozzi et al. 2004 for more details).   

\section{Predictions in the Herschel Bands of PACS}
By using our model and appropriate SEDs for the different source populations
(Arp220 for the ``starburst'' component, M51 for the ``normal'', the AGN1 SED
from Elvis et al. (1994) for type 1, Circinus for type 2 AGN), we have predicted
the expected source counts and redshift distributions in the Herschel bands
covered by the PACS instrument (75, 110 and 170 $\mu$m). Here we report only
the results for the 75 and 170-$\mu$m bands (see figure \ref{fig_cnt}).
For the ``starburst'' population we have considered different SEDs (M82, NGC6090
and Arp220), but we have found that a far-infrared bright galaxy template
as Arp220 was needed in order to reproduce the FIRBACK observed counts (Dole et al. 2001).
Using such template SED we find that PACS will reach the confusion limit
at 75 and 170 $\mu$m at 0.73 and 19.7 mJy respectively. At these flux density levels
about all the IR cosmic background will be resolved into discrete sources at 75 $\mu$m
(CIB($S_{75}>0.73$ mJy) = 10 nWm$^{-2}$sr$^{-1}$) and about 58\% at 170 $\mu$m 
(CIB($S_{170}>19.7$ mJy) = 8 nWm$^{-2}$sr$^{-1}$).
The number of sources expected at these limits are 21848 per deg$^{-2}$ and 3488
 per deg$^{-2}$ at 75 and 170 $\mu$m respectively. The great majority ($\sim$95\%) 
of these sources will be ``starburst'' galaxies peaking at redshift 1.


\begin{thebibliography}{}
\bibitem[Dole et~al.(2001)]{dole01} Dole H. et~al. 2001, A\&A 372, 702
\bibitem[Elbaz et~al.(1999)]{elbaz99} Elbaz D. et~al. 1999, A\&A 361, L37
\bibitem[Elvis et~al. (1994)]{elvis04} Elvis M. et~al., 1994, ApJS 95, 1
\bibitem[Gruppioni et~al.(2002)]{cg02} Gruppioni C. et~al., 2002, MNRAS 335, 831
\bibitem[Hauser \& Dwek(2001)]{hauser01} Hauser M.G. \& Dwek E., 2001, ARA\&A 39, 249
\bibitem[La Franca et~al.(2004)]{lafranca04} La Franca F. et~al., 2004, AJ 127, 3075
\bibitem[Lari et~al.(2001)]{lari02} Lari C. et~al., 2001, MNRAS 325, 1173
\bibitem[Marleau et~al.(2004)]{marl04} Marleau F. et~al., 2004, ApJS 154, 66
\bibitem[Papovich et~al.(2004)]{papovich04} Papovich C. et~al. 2004, ApJS 154, 70
\bibitem[Pozzi et~al.(2004)]{pozzi04} Pozzi F. et~al. 2004, ApJ 609, 122
\bibitem[Rowan-Robinson et~al.(2004)]{mrr04} Rowan-Robinson M. et~al., 2004, MNRAS 351, 1290
\end{thebibliography}
\end{document}